\documentclass{aa}  
\usepackage{graphicx}
\usepackage{txfonts}
\def\tmap{\textsc{\bfseries tmap}}
\def\pro2{\textsc{\bfseries pro2}}
\def\lte2{\textsc{\bfseries lte2}}
\def\atoms2{\textsc{\bfseries atoms2}}
\def\setf2{\textsc{\bfseries setf2}}
\def\line1prof{\mbox{\textsc{\bfseries line1}\raisebox{-0.5ex}{\bfseries--}\textsc{\bfseries prof}}}

\def\etal{{et\,al.}\ }
\def\elf{PG\,1159$-$035}
\newcommand{\Teff}{$T\mathrm{\hspace*{-0.4ex}_{eff}}$}
\newcommand{\logg}{$\log\,g$\hspace*{0.5ex}}

\begin{document}
\title{High-resolution ultraviolet spectroscopy of \elf\ with HST and
FUSE\thanks{Based  on observations with the NASA/ESA Hubble Space Telescope,
obtained at the Space Telescope Science  Institute, which is operated by the
Association of Universities for Research in Astronomy, Inc., under  NASA
contract NAS5-26666. Based on observations made with the NASA-CNES-CSA Far
Ultraviolet Spectroscopic  Explorer. FUSE is operated for NASA by the Johns
Hopkins University under NASA contract NAS5-32985.}}

\author{D\@. Jahn\inst{1}
   \and T\@. Rauch\inst{1}
   \and E\@. Reiff\inst{1}
   \and K\@. Werner\inst{1}
   \and J.W\@. Kruk\inst{2}
   \and F\@. Herwig\inst{3}}

\institute{Institut f\"ur Astronomie und Astrophysik, Universit\"at T\"ubingen, Sand 1, 72076 T\"ubingen, Germany
   \and Department of Physics and Astronomy, Johns Hopkins University, Baltimore, MD 21218, USA
   \and Los Alamos National Laboratory, Theoretical Astrophysics Group T-6, MS B227, Los Alamos, NM 87545, USA} 

\offprints{K\@. Werner}

\mail{werner@astro.uni-tuebingen.de}

\date{Received; accepted}

\authorrunning{D. Jahn et al.}
\titlerunning{HST and FUSE spectroscopy of \elf}

\abstract
{\elf\ is the prototype of the PG1159 spectral class which consists
of extremely hot hydrogen-deficient (pre-) white dwarfs. It is also the prototype of
the GW Vir variables, which are non-radial g-mode pulsators. The study of PG1159
stars reveals insight into stellar evolution and nucleosynthesis during AGB and
post-AGB phases.}
{We perform a quantitative spectral analysis of \elf\ focusing on the abundance
  determination of trace elements.}
{We have taken high-resolution ultraviolet spectra of \elf\ with the
  \emph{Hubble Space Telescope} and the \emph{Far Ultraviolet Spectroscopic
  Explorer}. They are analysed with non-LTE line blanketed model atmospheres.}
{We confirm the high effective temperature with high precision
  (\Teff=140\,000$\pm$5000~K) and the surface gravity of \logg=7. For the first
  time we assess the abundances of silicon, phosphorus, sulfur,
  and iron. Silicon is about solar. For phosphorus we find an upper limit of
  solar abundance. A surprisingly strong depletion of sulfur (2\% solar) is discovered. Iron
  is not detected, suggesting an upper limit of 30\% solar. This coincides
  with the Fe deficiency found in other PG1159 stars. We redetermine the
  nitrogen abundance and find it to be lower by one dex compared to previous
  analyses.}
{The sulfur depletion is in contradiction with current models of AGB star
  intershell nucleosynthesis. The iron deficiency confirms similar results for
  other PG1159 stars and is explained by the conversion of iron into heavier
  elements by n-capture in the s-processing environment of the precursor AGB
  star. However, the extent of the iron depletion is stronger than predicted by
  evolutionary models. The relatively low nitrogen abundance compared to other
  pulsating PG1159 stars weakens the role of 
nitrogen as a distinctive feature of pulsators and non-pulsators in the GW~Vir instability strip. }


\maketitle
%

\begin{table}
\begin{center}
\caption{Identified photospheric lines (rest wavelengths) in the FUSE spectrum
  of \elf.}
\label{tab:lines-FUSE} 
\begin{tabular}{rlrlrcl}  
\hline 
\hline 
\noalign{\smallskip}
\multicolumn{3}{c}{Wavelength / \AA } &   Ion & \multicolumn{3}{c}{Transition}  \\ \hline 
\noalign{\smallskip}
 933.38 &   &        &   \ion{S}{vi}  & $\mathrm{3s\ ^2S_{1/2}}$ &--& $\mathrm{3p\ ^2P^o_{3/2}}$ \\
 942.51 &   &        &   \ion{He}{ii} & 2 &--& 11       \\ 
 944.52 &   &        &   \ion{S}{vi}  & $\mathrm{3s\ ^2S_{1/2}}$ &--& $\mathrm{3p\ ^2P^o_{1/2}}$ \\ 
 948.09 &-- & 948.21 &   \ion{C}{iv}  & $\mathrm{3s\ ^2S}$ &--& $\mathrm{4p\ ^2P^o}$ \\ 
 949.33 &   &        &   \ion{He}{ii} & 2                         &--& 10       \\ 
 958.70 &   &        &   \ion{He}{ii} & 2                         &--& 9        \\ 
 966.60 &-- & 967.34 &   \ion{Si}{v}  & $\mathrm{3p\ ^3D}$ &--&$\mathrm{3d\ ^3F^o}$  \\ 
 967.19 &-- &        &   \ion{Si}{v}  & $\mathrm{3p\ ^1P}$ &--&$\mathrm{3d\ ^1D^o}$  \\ 
 970.45 &-- &970.61  &   \ion{O}{vi}  & $\mathrm{5s\ ^2S}$ &--& $\mathrm{8p\ ^2P^o}$  \\ 
 972.11 &   &        &   \ion{He}{ii} & 2      &--& 8        \\ 
 973.33 &   &        &   \ion{Ne}{vii}& $\mathrm{2p\ ^1P^o}$ &--&$\mathrm{2p^{2\ 1}D}$  \\ 
 975.83 &   &        &   \ion{Si}{v}  & $\mathrm{3p\ ^3P}$ &--&$\mathrm{3d\ ^3D^o}$  \\ 
 986.32 &-- & 986.38 &   \ion{O}{vi}  & $\mathrm{4s\ ^2S}$ &--& $\mathrm{5p\ ^2P^o}$  \\ 
 987.81 &   &        &   \ion{Si}{v}  & $\mathrm{3p\ ^1D}$ &--&$\mathrm{3d\ ^1F^o}$  \\ 
 989.47 &   &        &   \ion{C}{iv}  & $\mathrm{4s\ ^2S}$ &--&  $\mathrm{10p\ ^2P^o}$  \\ 
 992.36 &   &        &   \ion{He}{ii} & 2      &--& 7        \\ 
1018.75 &-- &1018.88 &   \ion{O}{vi}  & $\mathrm{5p\ ^2P^o}$&--&$\mathrm{8d\ ^2D}$  \\ 
1025.27 &   &        &   \ion{He}{ii} & 2      &--& 6        \\ 
1031.27 &-- &1031.28 &   \ion{C}{iv}  & $\mathrm{4s\ ^2S}$ &--&  $\mathrm{9p\ ^2P^o}$  \\ 
1031.91 &   &        &   \ion{O}{vi}  & $\mathrm{2s\ ^2 S}$ &--& $\mathrm{2p\ ^2P^o}$  \\ 
1037.61 &   &        &   \ion{O}{vi}  & $\mathrm{2s\ ^2 S}$ &--& $\mathrm{2p\ ^2P^o}$  \\ 
1038.47 &   &        &   \ion{O}{vi}  & $\mathrm{5g\ ^2G}$&--& $\mathrm{8h\ ^2H^o}$   \\
1050.   &   &        &   \ion{C}{iv}  & 4  &--& 11\\               
1060.59 &-- &1060.74 &   \ion{C}{iv}  & $\mathrm{4p\ ^2P^o}$ &--& $\mathrm{10d\ ^2D}$  \\ 
1066.63 &-- &1066.78 &   \ion{C}{iv}  & $\mathrm{4p\ ^2P^o}$&--& $\mathrm{10s\ ^2S}$    \\ 
1080.60 &-- &1081.34 &   \ion{O}{vi}  & $\mathrm{4p\ ^2P^o}$&--& $\mathrm{5d\ ^2D}$   \\ 
1083.62 &-- &1083.67 &   \ion{C}{iv}  & $\mathrm{4d\ ^2D}$ &--& $\mathrm{10f\ ^2F^o}$ \\ 
1084.77 &   &        &   \ion{C}{iv}  & $\mathrm{4f\ ^2F^o}$ &--& $\mathrm{10g\ ^2G}$ \\ 
1084.94 &   &        &   \ion{He}{ii} & 2         &--& 5        \\ 
1097.32 &-- &1097.34 &   \ion{C}{iv}  & $\mathrm{4s\ ^2 S}$&--& $\mathrm{8p\ ^2P^o}$  \\ 
1107.59 &-- &1107.98 &   \ion{C}{iv}  & $\mathrm{3p\ ^2P^o}$&--& $\mathrm{4d\ ^2D}$   \\ 
1108.89 &-- &1109.06 &   \ion{C}{iv}  & $\mathrm{4p\ ^2P^o}$&--& $\mathrm{9d\ ^2D}$   \\ 
1118.81 &   &        &   \ion{Si}{v}  & $\mathrm{3s\ ^3P^o}$ &--&$\mathrm{3p\ ^3P}$  \\ 
1122.33 &-- &1122.61 &   \ion{O}{vi}  & $\mathrm{4d\ ^2D}$ &--& $\mathrm{5f\ ^2F^o}$  \\ 
1124.70 &-- &1124.82 &   \ion{O}{vi}  & $\mathrm{4f\ ^2F^o}$&--& $\mathrm{5g\ ^2G}$  \\ 
1126.17 &-- &1126.28 &   \ion{O}{vi}  & $\mathrm{4f\ ^2F^o}$&--& $\mathrm{5d\ ^2D}$   \\ 
1134.25 &-- &1134.30 &   \ion{C}{iv}  & $\mathrm{4d\ ^2D}$ &--& $\mathrm{9f\ ^2F^o}$  \\ 
1135.50 &   &        &   \ion{C}{iv}  & $\mathrm{4f\ ^2F^o}$&--& $\mathrm{9g\ ^2G}$   \\ 
1135.64 &   &        &   \ion{C}{iv}  & $\mathrm{4f\ ^2F^o}$&--& $\mathrm{9d\ ^2D}$  \\ 
1136.63 &-- &1136.68 &   \ion{C}{iv}  & $\mathrm{4d\ ^2D}$ &--&  $\mathrm{9p\ ^2P^o}$ \\ 
1139.50 &   &        &   \ion{F}{vi}  & $\mathrm{2s2p\  ^1P^o}$&--&$\mathrm{2p^{2\ 1} D}$  \\ 
1146.75 &-- &1147.02 &   \ion{O}{vi}  & $\mathrm{4d\ ^2D}$ &--& $\mathrm{5p\ ^2P^o}$  \\ 
1168.85 &-- &1168.99 &   \ion{C}{iv}  & $\mathrm{3d\ ^2D}$ &--& $\mathrm{4f\ ^2F^o}$  \\
1171.12 &-- &1172.00 &   \ion{O}{vi}  & $\mathrm{4p\ ^2P^o}$&--& $\mathrm{5s\ ^2 S}$  \\  
1184.59 &-- &1184.77 &   \ion{C}{iv}  & $\mathrm{4p\ ^2P^o}$&--&  $\mathrm{8d\ ^2D}$  \\ 
\noalign{\smallskip} \hline
\end{tabular} 
\end{center}
\end{table}

\begin{table}
\begin{center}
\caption{Photospheric lines (rest wavelengths) in the HST spectrum.}
\label{tab:lines-STIS} 
\begin{tabular}{rlrlrcl}
\hline 
\hline 
\noalign{\smallskip}
\multicolumn{3}{c}{Wavelength / \AA } &  Ion & \multicolumn{3}{c}{Transition} \\ \hline
\noalign{\smallskip} 
1168.84 &-- & 1169.99  &  \ion{C}{iv}  &$\mathrm{3d\ ^2D}$  &--& $\mathrm{4f\ ^2F^o}$    \\
1171.12 &-- & 1172.00  &  \ion{O}{vi}  &$\mathrm{4p\ ^2P^o}$ &--& $\mathrm{5s\ ^2 S}$    \\
1184.58 &-- & 1184.77  &  \ion{C}{iv}  &$\mathrm{4p\ ^2P^o}$ &--&$\mathrm{8d\ ^2D}$    \\
1198.40 &-- & 1198.59  &  \ion{C}{iv}  &$\mathrm{3d\ ^2D}$  &--& $\mathrm{4p\ ^2P^o}$    \\
1199.91 &-- & 1200.10  &  \ion{C}{iv}  &$\mathrm{4p\ ^2P^o}$ &--&$\mathrm{8s\ ^2 S}$     \\
1207.69 &   &          &  \ion{Si}{vi} &$\mathrm{3s\ ^2D}$ &--&$\mathrm{3p\ ^2F^o}$     \\
1210.61 &-- & 1210.65  &  \ion{C}{iv}  & $\mathrm{4s\ ^2 S}$ &--& $\mathrm{7p\ ^2P^o}$   \\
1218.50 &   &          &  \ion{Si}{vi} &$\mathrm{3s\ ^2D}$ &--&$\mathrm{3p\ ^2F^o}$     \\
1229.01 &   &          &  \ion{Si}{vi} &$\mathrm{3s\ ^2P}$ &--&$\mathrm{3p\ ^2D^o}$     \\
1230.04 &-- & 1230.52  &  \ion{C}{iv}  & $\mathrm{3p\ ^2P^o}$ &--& $\mathrm{4s\ ^2 S}$    \\
1235.45 &   &          &  \ion{Si}{v}  & $\mathrm{3s\ ^3P^o}$  &--& $\mathrm{3p\ ^3D}$   \\
1238.82 &-- & 1242.80  &  \ion{N}{v}   & $\mathrm{2s\ ^2 S}$  &--& $\mathrm{2p\ ^2P^o}$   \\
1243.33 &   &          &  \ion{Si}{vi} &$\mathrm{3s\ ^2P}$ &--&$\mathrm{3p\ ^2D^o}$     \\
1245.73 &   &          &  \ion{Si}{v}  & $\mathrm{3s\ ^3P^o}$  &--& $\mathrm{3p\ ^3D}$   \\
1251.39 &   &          &  \ion{Si}{v}  & $\mathrm{3s\ ^3P^o}$  &--& $\mathrm{3p\ ^3D}$   \\
1261.42 &-- & 1261.80  &  \ion{O}{vi}  & $\mathrm{5p\ ^2P^o}$ &--& $\mathrm{7d\ ^2D}$     \\
1276.01 &   &          &  \ion{Si}{v}  & $\mathrm{3s\ ^3P^o}$  &--& $\mathrm{3p\ ^3D}$   \\
1290.06 &-- & 1290.21  &  \ion{O}{vi}  & $\mathrm{5d\ ^2D}$  &--& $\mathrm{7f\ ^2F^o}$   \\
1291.81 &-- & 1291.84  &  \ion{O}{vi}  & $\mathrm{5f\ ^2F^o}$ &--& $\mathrm{7g\ ^2G}$    \\
1291.89 &-- & 1291.92  &  \ion{O}{vi}  & $\mathrm{5g\ ^2G}$ &--& $\mathrm{7h\ ^2H^o}$    \\
1291.99 &-- & 1292.02  &  \ion{O}{vi}  & $\mathrm{5g\ ^2G}$ &--& $\mathrm{7f\ ^2F^o}$    \\
1292.64 &-- & 1293.04  &  \ion{O}{vi}  & $\mathrm{5f\ ^2F^o}$ &--& $\mathrm{7d\ ^2D}$     \\
1298.   &   &          &  \ion{O}{vi}  & $\mathrm{6}$ &--& $\mathrm{11}$     \\
1315.62 &-- & 1315.86  &  \ion{C}{iv}  & $\mathrm{4p\ ^2P^o}$ &--& $\mathrm{7d\ ^2D}$    \\
1319.60 &   &          &  \ion{Si}{v}  & $\mathrm{3s\ ^1P^o}$  &--& $\mathrm{3p\ ^1D}$   \\
1319.71 &   &          &  \ion{Ne}{vii}& $\mathrm{2p\ ^1P^o}$ &--& $\mathrm{2p^{2\ 3}P}$  \\
1344.18 &-- & 1344.41  &  \ion{C}{iv}  & $\mathrm{4p\ ^2P^o}$ &--& $\mathrm{7s\ ^2 S}$      \\
1351.21 &-- & 1351.29  &  \ion{C}{iv}  & $\mathrm{4d\ ^2D}$  &--& $\mathrm{7f\ ^2F^o}$    \\
1352.97 &-- & 1352.97  &  \ion{C}{iv}  & $\mathrm{4f\ ^2F^o}$ &--& $\mathrm{7g\ ^2G}$    \\
1353.43 &   &          &  \ion{C}{iv}  & $\mathrm{4f\ ^2F^o}$ &--& $\mathrm{7d\ ^2D}$    \\
1358.42 &-- & 1358.50  &  \ion{C}{iv}  & $\mathrm{4d\ ^2D}$  &--& $\mathrm{7p\ ^2P^o}$    \\
1371.30 &  &           &  \ion{O}{v}   &$\mathrm{2p\ ^1P^o}$ &--& $\mathrm{2p^2\ ^1D}$   \\
1393.75 &-- & 1402.77   &  \ion{Si}{iv} &$\mathrm{3s\ ^2 S}$ &--&$\mathrm{3p\ ^2P^o}$    \\
1422.65 &--  & 1422.81 &  \ion{O}{vi} &$\mathrm{6d\ ^2 D}$ &--&$\mathrm{10f\ ^2F^o}$    \\
1423.61 &--  & 1423.65  &  \ion{O}{vi} &$\mathrm{6g\ ^2 G}$ &--&$\mathrm{10h\ ^2H^o}$    \\
1423.74 &--  & 1423.79  &  \ion{O}{vi} &$\mathrm{6g\ ^2 G}$ &--&$\mathrm{10f\ ^2F^o}$    \\
1423.83 &--  & 1423.97  &  \ion{O}{vi} &$\mathrm{6p\ ^2P^o}$ &--& $\mathrm{10s\ ^2 S}$  \\
1423.90 &--  &1423.92  &  \ion{O}{vi} &$\mathrm{6h\ ^2 H^o}$ &--&$\mathrm{10i\ ^2I}$    \\ 
1423.92 &--  & 1423.94 &  \ion{O}{vi} &$\mathrm{6h\ ^2 H^o}$ &--&$\mathrm{10g\ ^2G}$    \\
1424.39 &--  & 1424.48 &  \ion{O}{vi} &$\mathrm{6f\ ^2F^o}$ &--&$\mathrm{10g\ ^2G}$    \\
1424.63 &--  &1424.72  &  \ion{O}{vi} &$\mathrm{6f\ ^2F^o}$ &--&$\mathrm{10d\ ^2D}$    \\
1427.40 &--  &1427.56  &  \ion{O}{vi} &$\mathrm{6d\ ^2 D}$ &--&$\mathrm{10p\ ^2P^o}$    \\
1440.28 &--  &1440.36  &  \ion{C}{iv}  &$\mathrm{4s\ ^2 S}$ &--&$\mathrm{6p\ ^2P^o}$   \\
1548.20 &--  &1550.77  &  \ion{C}{iv}  & $\mathrm{2s\ ^2 S}$  &--& $\mathrm{2p\ ^2P^o}$    \\
1585.81 &-- & 1586.14  &  \ion{C}{iv}  & $\mathrm{4p\ ^2P^o}$ &--&  $\mathrm{6d\ ^2D}$   \\ 
1637.54 &-- & 1637.65  &  \ion{C}{iv}  & $\mathrm{4d\ ^2D}$  &--& $\mathrm{6f\ ^2F^o}$   \\
1638.73 &-- & 1638.94  &  \ion{O}{vi}  &  $\mathrm{6d\ ^2D}$  &--& $\mathrm{9f\ ^2F^o}$   \\
1639.85 &-- &  1639.91 &  \ion{O}{vi}  & $\mathrm{6g\ ^2 G}$ &--&$\mathrm{9h\ ^2H^o}$    \\ 
1640.10&   &          &  \ion{C}{iv}  & $\mathrm{4f\ ^2F^o}$ &--& $\mathrm{6g\ ^2G}$    \\
1640.18 &-- & 1640.24 &  \ion{O}{vi}  & $\mathrm{6g\ ^2 G}$ &--&$\mathrm{9f\ ^2F^o}$    \\
1640.33 &-- & 1640.36 &  \ion{O}{vi}  & $\mathrm{6h\ ^2 H^o}$ &--&$\mathrm{9i\ ^2I}$    \\ 
1640.38 &-- & 1640.41 &  \ion{O}{vi}  & $\mathrm{6h\ ^2 H^o}$ &--&$\mathrm{9g\ ^2G}$    \\ 
1640.42&   &          &  \ion{He}{ii}  & 2 &--& 3    \\
1640.98 &-- & 1641.10 &  \ion{O}{vi}  & $\mathrm{6f\ ^2F^o}$ &--&$\mathrm{9g\ ^2G}$    \\
1641.45 &-- & 1641.57 &  \ion{O}{vi}  & $\mathrm{6f\ ^2F^o}$ &--&$\mathrm{9d\ ^2D}$    \\
1647.35 &-- &  1647.56&  \ion{O}{vi}  & $\mathrm{6d\ ^2 D}$ &--&$\mathrm{9p\ ^2P^o}$    \\
1649.50 &-- &  1649.68&  \ion{O}{vi}  & $\mathrm{6p\ ^2P^o}$ &--& $\mathrm{9s\ ^2 S}$  \\
1653.63&-- &1653.99   &  \ion{C}{iv}  & $\mathrm{4p\ ^2P^o}$ &--& $\mathrm{6s\ ^2S}$    \\
1654.46&-- & 1654.57  &  \ion{C}{iv}  & $\mathrm{4d\ ^2D}$ &--& $\mathrm{6p\ ^2P^o}$    \\ 
\noalign{\smallskip} \hline
\end{tabular} 
\end{center}
\end{table}

\section{Introduction}
\label{intro}

\elf\ is the prototype star of the PG1159 spectral class, which is a group of 40
extremely hot (\Teff=75\,000--200\,000~K) hydrogen-deficient post-AGB
stars. They represent the evolutionary transition phase between Wolf-Rayet type
central stars of planetary nebulae and the hottest non-DA white dwarfs. Their
H-deficiency is probably caused by a late He-shell flash, which has laid bare
the intershell region of the precursor red giant (see, e.g., a recent review by
Werner \& Herwig 2006). The study of the chemical composition of PG1159 stars is
therefore of particular interest, because it allows us to directly analyse the
outcome of mixing processes and nucleosynthesis in AGB star interiors.

\elf\ (=GW~Vir) is also the prototype of variable objects among the PG1159
stars, which also comprises a number of [WC] central stars. They are non-radial
g-mode pulsators. Besides the Sun, \elf\ is probably the star most intensively
studied by asteroseismic methods (e.g.\ Kawaler \& Bradley 1994). Exploration of
the interior structure allows insight into previous AGB evolution stages and is
thus of significance well beyond the PG1159 stars themselves. The basic
pulsation driving mechanism is the $\kappa-\gamma$ effect of C and O acting
beneath the photosphere (Starrfield \etal 1983), however, many details
concerning the role of abundances of other species are still debated (e.g.,
Gautschy \etal 2005).

The particular significance of \elf\ for insight into (post-) AGB  stellar
structure and evolution motivates our work to perform a detailed spectroscopic
analysis of this star. We have taken high-resolution UV spectra with the
\emph{Far Ultraviolet Spectroscopic Explorer} (FUSE) and the \emph{Hubble Space
Telescope} (HST) that cover the wavelength interval from the Lyman edge up to
1730~\AA. The emphasis of our analysis is on the identification and
abundance analysis of trace elements. The results will be discussed in the
framework of late He-shell flash and AGB evolution models as well as pulsation
theory.

\elf\ was discovered in the  Palomar Green Survey (Green \etal 1986, Wesemael
\etal 1985) and was first subject to a quantitative non-LTE spectral analysis by
Werner \etal (1991). Based on optical spectra, \Teff=140\,000~K and \logg=7.0
were derived and the main atmospheric constituents were found to be He=33\%,
C=50\%, and O=17\% (mass fractions). Later work including the present one
confirmed these parameters and we concentrate here on other chemical elements.

UV spectroscopy has been performed with the \emph{International
Ultraviolet Explorer} (IUE, Liebert \etal 1989), the \emph{Hopkins Ultraviolet
Telescope} (HUT, Kruk \& Werner 1998), as well as with the \emph{Faint Object
Spectrograph} aboard HST (Werner \& Heber 1993, Dreizler \& Heber 1998). It
revealed that, as in the optical band, the spectra are dominated by lines from
\ion{He}{ii}, \ion{C}{iv} and \ion{O}{vi}.  The resolution and S/N of the
spectra presented here are far superior. We will refer to results from earlier
observations where relevant.

The paper is organized as follows. We present our new HST and FUSE
observations in Sect.\,\ref{obs} and identify photospheric lines. In
Sect.\,\ref{model} we describe our model atmosphere and line formation
calculations. Sect.~\ref{analysis} contains the results of the spectral analysis
and we conclude with a discussion of our results in Sect.\,\ref{discussion}.

\begin{figure*}
\begin{center}
\includegraphics[width=0.97\textwidth]{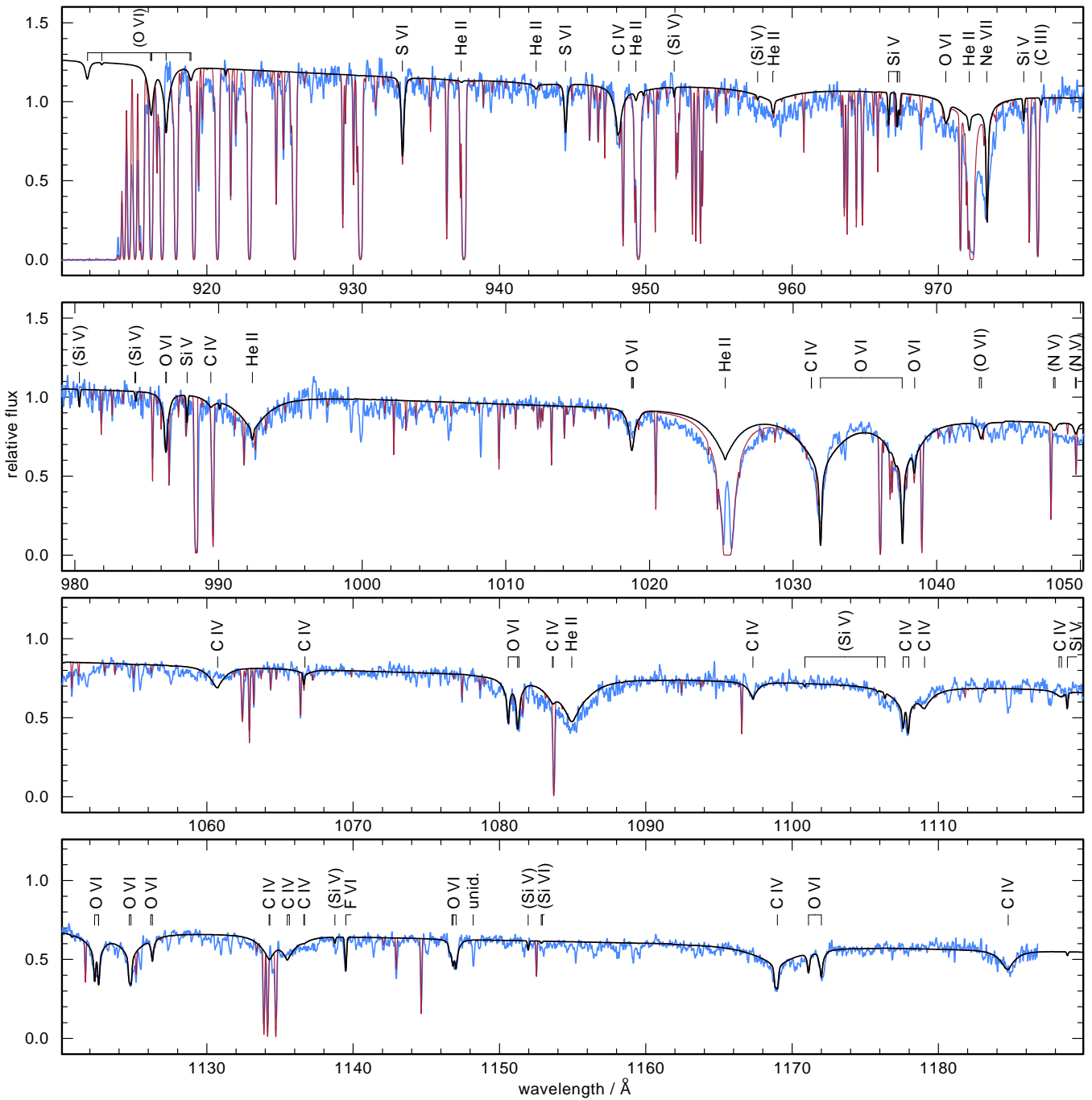}
\caption{The complete FUSE spectrum of \elf\ compared to our final model
  spectrum (black graph). The
  detected photospheric lines are identified. Labels in brackets denote lines
  that are visible in the model but not unambiguously seen in the
  observation. Overplotted is the same model but now attenuated by ISM absorption
  lines (grey graph; in the online
  version of this paper the graph is red). 
The ISM lines are not labeled. Photospheric model parameters are given in
  Table~\ref{tab:results}. In all figures of this paper the observed spectra are
  shifted such that the photospheric lines are located at rest wavelengths.} 
\label{fig:FUSE} 
\end{center}
\end{figure*}  

\begin{figure*}
\begin{center}
\includegraphics[width=0.95\textwidth]{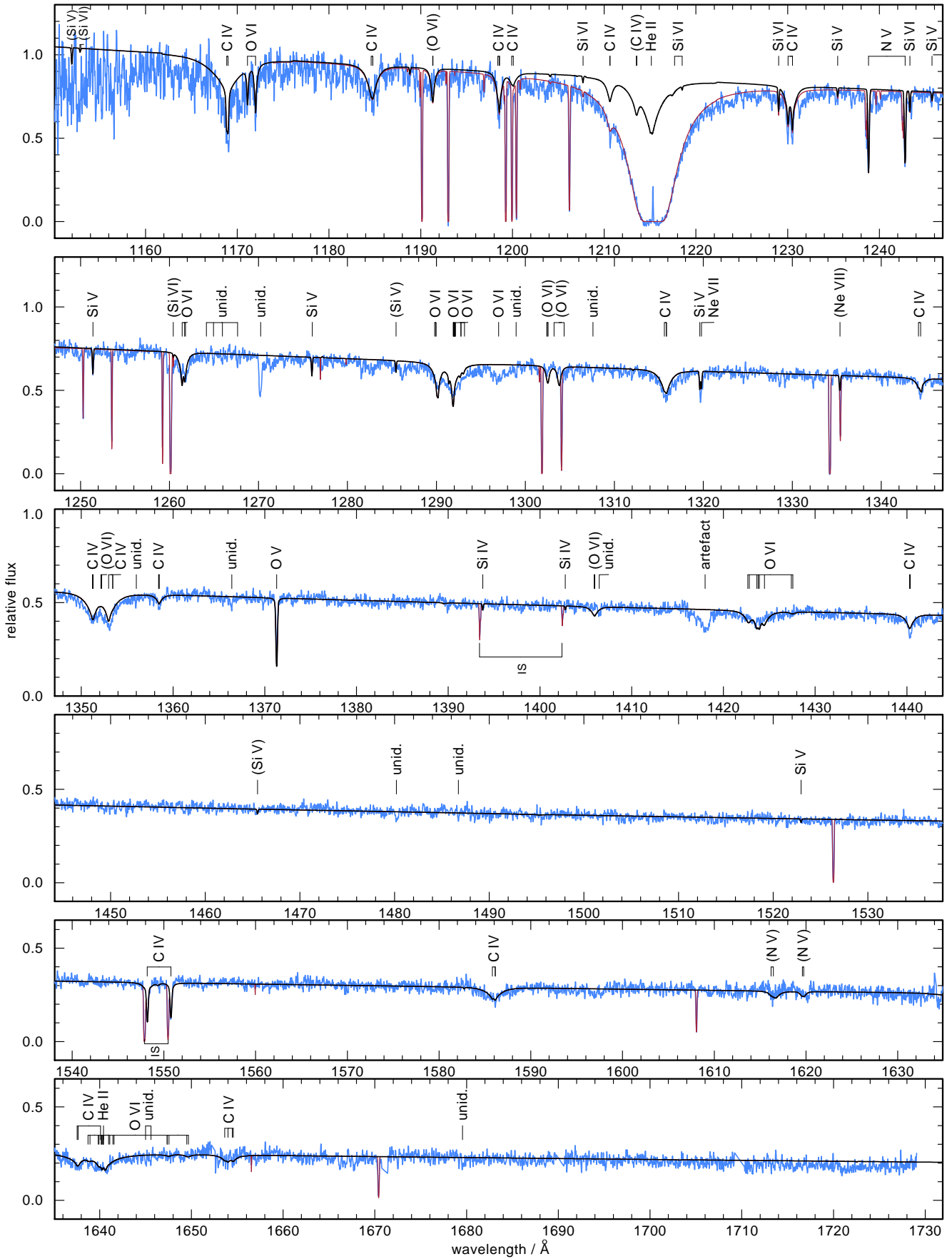}
\caption{Similar to Fig.\,\ref{fig:FUSE}. Here we show the complete HST spectrum
  of \elf. In addition to the photospheric lines we also label the interstellar
  components of the resonance doublets from \ion{C}{iv} and \ion{Si}{iv}.   } 
\label{fig:STIS} 
\end{center}
\end{figure*}  

\begin{table}
\begin{center}
\caption{Radial-velocity shift of selected photospheric lines.}
\label{tab:radvel}
\begin{tabular}{ccccc}
\hline
\hline
\noalign{\smallskip}
 Ion & $\lambda_\mathrm{rest}$ / \AA & $\lambda_\mathrm{observed}$ / \AA& $v_\mathrm{rad}$ / km/s \\
\noalign{\smallskip} \hline
\noalign{\smallskip}
\ion{N}{v}   &1238.82  &1239.06  & 58.08 \\
\ion{N}{v}   &1242.80  &1243.04  & 57.89 \\
\ion{Si}{iv} &1393.76  &1394.07  & 66.68 \\
\ion{C}{iv}  &1548.19  &1548.48  & 56.16 \\
\ion{C}{iv}  &1550.77  &1551.06  & 56.06 \\
\ion{O}{v}   &1371.30  &1371.57  & 59.90 \\
\noalign{\smallskip} \hline
\noalign{\smallskip} 
Mean         &         &         & 59.13 $\pm 1.62$\\
\noalign{\smallskip} \hline
\end{tabular} \\
\end{center}
\end{table}

\section{Observations}
\label{obs}

\subsection{FUSE and HST spectroscopy}

The FUSE spectrum (observation id Q1090101) was obtained on May 10, 2001,  using
the LWRS aperture. The observation was split evenly into three  exposures in
successive orbits for a total exposure time of 6324~sec.  The spectrum covers
the range from the Lyman edge up to 1187~\AA, i.e., there is a narrow range
that overlaps with the HST data.  The data were reduced with version 3.1.7 of
the CalFUSE pipeline.   The spectral resolution was rather poor in this first
reduction, about 10\,000, corresponding to $\approx$0.1~\AA, indicating that
motion of the optics during the exposure might have been significant (Sahnow
\etal 2000).  The exposures were then reprocessed in segments of 400-500 sec
each.  The spectra from these separate sub-exposures were coaligned  by
cross-correlation on narrow interstellar absorption lines, and combined.  The
resulting spectral resolution was much closer to the expected value of 20\,000.

The spectra from the four channels were then combined to produce a single
spectrum.  The spectra were first resampled onto a common wavelength scale: the
pixel width of 0.013~\AA\ was not changed, but non-integer pixel shifts  were
required.  This causes a slight smoothing of the spectra, but the effect is
small in comparison with the instrumental resolution width.  At some wavelengths
additional shifts of 1 to 2 pixels were required to coalign the spectra from
different channels, as a result of small residual distortions in the detector
pixel scale.  Regions of known bad pixels were discarded.  The broad depression
in the LiF\,1b spectrum induced by the grid wires in the detector was corrected
by normalizing the LiF\,1b spectrum to LiF\,2a with a 100-pixel sliding boxcar
filter.  A similar but weaker depression in the SiC\,1b flux over wavelengths of
977--992~\AA\ was corrected by normalizing to the SiC\,2a spectrum in the same
manner.  The spectra were weighted by the statistical uncertainties on a
pixel-by-pixel basis when combined to produce the final spectrum, which is
displayed in Fig.\,\ref{fig:FUSE}.

The HST spectrum was obtained on June 9, 2004, with the \emph{Space Telescope
Imaging Spectrograph} (STIS) using the grating E140M and the
$0.2\arcsec\times0.2\arcsec$ aperture. The exposure time was 2275~sec, divided
evenly among three exposures. The spectrum covers the range 1150--1730~\AA\ with
a resolution of 90\,000 corresponding to roughly 0.02~\AA. The standard
calibrated data files provided by the Multimission Archive at Space Telescope
(MAST) were used, which were reduced using the most recent version of the
CalSTIS pipeline (v 2.20) and associated reference files.  These spectra were
resampled onto a common wavelength scale and combined, weighting by the
statistical uncertainties of the data.  The pixel width of the wavelength scale
varied linearly from 0.0123~\AA\ at 1140~\AA\ to 0.01826~\AA\ at 1729~\AA,
closely approximating the sampling of the individual echelle orders.  Pixels at
the ends of the echelle orders with bad quality flags were discarded.  The final
merged spectrum is displayed  in Fig.\,\ref{fig:STIS}.

The FUSE and STIS spectra in Figs.\,\ref{fig:FUSE} and \ref{fig:STIS} are
overplotted with the astrophysical flux (cgs\AA\ units) from our final model,
normalized to the local continuum.

\subsection{Line identification}

Line identification was guided by model-atmosphere spectra that involve atomic
data from different sources, which are described in detail in
Sect.\,\ref{atomic_data}.

Table~\ref{tab:lines-FUSE} lists all photospheric lines detected in the FUSE
spectrum. The majority stems from \ion{He}{ii}, \ion{C}{iv}, and \ion{O}{vi} and
was already detected in the HUT spectrum (Kruk \& Werner 1998). However, thanks
to its higher resolution the FUSE spectrum allows for the identification of
individual multiplet components of several \ion{C}{iv} and \ion{O}{vi}
lines. The discovery of highly ionized neon and fluorine lines
(\ion{Ne}{vii}~973.33~\AA, \ion{F}{vi}~1139.50~\AA) in the FUSE spectra of \elf\
and other PG1159 stars was announced recently (Werner \etal 2004, 2005). In
addition, we present here the discovery of the \ion{S}{vi}~933.38/944.52~\AA\
resonance doublet. We also identify for the first time several lines from
\ion{Si}{v}; we discuss this further below together with other silicon features
(\ion{Si}{iv-vi}) seen in the HST spectrum.

We cannot confirm our earlier claimed detection of \ion{N}{v} lines in the HUT
spectrum. A broad and shallow feature at 1050~\AA\ was attributed to the 4d--6f
and 4f--6g transitions at 1048.3 and 1049.7~\AA, which was supported by a model
spectrum calculated with 1\% nitrogen. In this paper we derive from the
\ion{N}{v} resonance doublet in the HST spectrum that the N abundance is a
factor of 10 lower. As a consequence, the two lines at 1048.3 and 1049.7~\AA\
are very weak in our final model (Fig.\,\ref{fig:FUSE}). The two narrow lines in
the FUSE spectrum close to the wavelengths of the \ion{N}{v} lines are
interstellar lines of \ion{Ar}{i}~1048.22~\AA\ and H$_2$ R(0)$_{4-0}$
1049.374~\AA.  As in  the HUT spectrum, the FUSE spectrum also shows a broad
feature at 1048-1052~\AA. An alternative identification could be a very highly
excited \ion{C}{iv} line (n=4--11) which, however, cannot be checked with our
models because the upper level is not treated in NLTE in our model atom.

The FUSE spectra show that the \ion{C}{iii}~977.0~\AA\ line detected in the HUT
spectrum is entirely of interstellar origin:  the Doppler shift matches the
velocity of the other interstellar absorption features, and the predicted line
feature in our model atmosphere spectrum is very weak.

We will later discuss the non-detection of the \ion{P}{v} 1117.98/1128.01~\AA\
resonance doublet. It was discovered in cooler PG1159 stars (Reiff \etal 2006)
but \elf\ is obviously too hot to exhibit these lines, so that we will only be
able to derive an upper abundance limit.

\begin{figure}
\begin{center}
\includegraphics[width=0.48\textwidth]{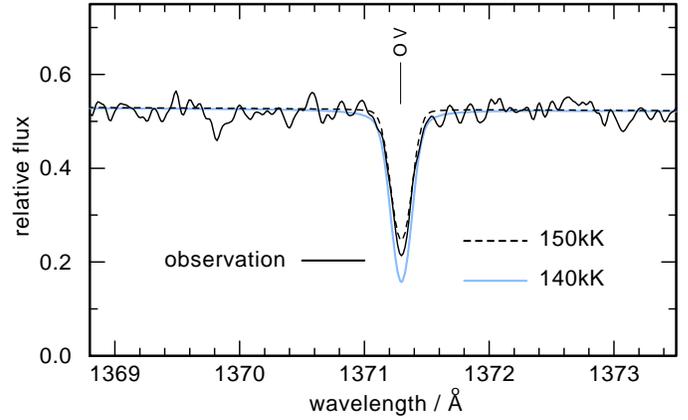}
\caption{The \ion{O}{v}~1371.30~\AA\ line compared with two models with different
  effective temperatures.} 
\label{fig:T-Vgl} 
\end{center}
\end{figure}  

Table~\ref{tab:lines-STIS}  lists all photospheric lines identified in the HST
spectrum. The majority stems from \ion{C}{iv} and \ion{O}{vi}. In addition we
see a broad \ion{He}{ii}~1640~\AA\ line, blended in a wide trough with
\ion{C}{iv} and \ion{O}{vi} lines (Fig.\,\ref{fig:STIS}).  The
\ion{O}{v}~1371~\AA\ line is seen with a narrow and deep profile
(Fig.\,\ref{fig:T-Vgl}). Concerning the \ion{N}{v}~1239/1243~\AA\ resonance
doublet, our spectrum allows for a clear distinction between interstellar and
photospheric components and, thus, for a reliable abundance determination.

We also discovered the \ion{Si}{iv}~1394/1403~\AA\ resonance doublet, but it is
very weak (Fig.\,\ref{fig:Si-Vgl}). Identification of the blue component is
rather doubtful although it should be stronger than the red one. We were able to
find stronger silicon lines from higher ionisation stages, \ion{Si}{v} and
\ion{Si}{vi}, in the FUSE and HST spectra (Fig.\,\ref{fig:Si-Vgl1}). To our
knowledge, this is the first detection of such lines in a stellar photospheric
spectrum. Accordingly, Fig.\,\ref{fig:ionfrac} shows that these two ionisation
stages are more strongly populated than \ion{Si}{iv} throughout the
atmosphere.

There is a quite strong absorption feature at 1319.7~\AA\ (lowest panel of
Fig.\,\ref{fig:Si-Vgl1}). It was discovered by Feibelman (1995) in an
IUE spectrum of the PG1159-type central star of NGC\,246. A close inspection of
the profile shows that the feature is a blend of two lines, separated by about
0.2~\AA. We suggest that they are lines from neon and silicon although our model
is neither able to fit the position accurately nor to match the strength of the
feature. One of the two lines could be a component of the \ion{Ne}{vii}
$\mathrm{2p ^1P^o}$--$\mathrm{2p^{2\ 3}P}$ intercombination triplet located at
1319.78/1335.35/1344.52~\AA. (Unfortunately, the two other components are
blended by a strong interstellar line and a \ion{C}{iv} line, preventing an
unambiguous identification.) The second line blending with this \ion{Ne}{vii}
component is probably \ion{Si}{v} $\mathrm{3s\ ^1P^o}$--$\mathrm{3p\ ^1D}$,
located at 1319.60~\AA.

We do not confirm the claimed detection of an \ion{O}{vii} line (1522~\AA) in
the IUE spectrum (Feibelman 1999).

Probably most other absorption lines in the HST spectrum are of interstellar
origin, however, there is a number of features that we think are photospheric
but which we are unable to identify (see Sect.\,\ref{unid_lines}).

From photospheric lines in the HST spectrum we have determined a radial velocity
shift of v$_{rad}=59.13\pm 1.62$~km/s (Tab.~\ref{tab:radvel}). Using the same
set of lines in IUE spectra, Holberg \etal (1998) determined v$_{rad}=50.10\pm
1.10$~km/s.  For the same set of ISM lines that was used by Holberg \etal (1998)
we derive v$_{rad}=-9.78\pm 0.59$~km/s as compared to their value of
v$_{rad}=-17.73\pm 1.06$~km/s. In all our Figures showing the HST spectrum we
have shifted the observation such that the photospheric lines are at their rest
wavelength position.

All observed spectra displayed here are convolved with Gaussians with
FWHM=0.05\AA\ and the synthetic spectra with FWHM=0.1\AA.

\begin{table*}
\begin{center}
\caption{Unidentified photospheric lines in the FUSE and HST spectra
(rest wavelengths). We remark whether the lines are detectable in
high-resolution spectra of other PG1159 stars or H-rich central stars as
well.}
\label{tab:unidentified photospheric lines} 
\begin{tabular}{ccc} 
\hline 
\hline 
\noalign{\smallskip}
wavelength / \AA  &   Remark & Possible identification\\ \hline
\noalign{\smallskip} 
927.30       &              & \\
939.57       &              & \\
965.35       &              & \\
996.58       &              & \\
999.42       &              & \\
1000.08       &              & \\
1006.20       &              & \\
1041--1047       &  some features            & \\
1052.00       &              & \\
1055.66	       &              & \\
1057.74       &              & \\
1110--1120       &  some features            & \\
1125.68       &              & \\
1127.57       &              & \\
1129.5        &              & \\
1131.17       &              & \\
1131.80       &              & \\
1139.04       &              & \\
1141.60       &              & \\
1148.42       &              & \\
1148--1152    & K1-16, RX\,J2117+3412, Lo4, PG1144+005, PG1707+427 & broad dip, \ion{C}{iv} 4--11\\
1155-1160       &  some features            & \\
1264.15, 1264.95, 1265.95 & NGC 7094, H-rich CSPN \\
1267.65       & NGC 7094, H-rich CSPN & 1267.64 \ion{Si}{v} $\mathrm{3s ^1P^o}$--$\mathrm{3p^{3}P}$, no f-value available \\
1270.25       &     NGC 246, NGC 7094, H-rich CSPN \\
1298.95       &    H-rich CSPN \\
1307.60       &   H-rich CSPN \\
1356.00       &   H-rich CSPN \\
1366.40       &   RX\,J2117.1+3412  (HST/GHRS, Werner \etal 1996)\\
1406.45       &   H-rich CSPN \\
1480.20       &   NGC 246, not in H-rich CSPN \\
1486.75       &  H-rich CSPN \\
1645.00, 1645.60   &   NGC 7094, H-rich CSPN       \\ 
1679.55       &  NGC 7094, H-rich CSPN \\
\hline
\end{tabular}
\end{center}
\end{table*}

\subsection{Unidentified photospheric lines}
\label{unid_lines}

In Table~\ref{tab:unidentified photospheric lines} we list unidentified lines
that we think are photospheric. We have carefully compared the FUSE and HST
spectra of \elf\ to spectra of other hot post-AGB stars whose photospheric and
ISM radial velocity systems have different separations. In the Table we note in
which objects we have seen the same feature as in \elf. Most useful for this
comparison were other HST/STIS spectra, namely that of the PG1159 star
NGC\,7094, which we observed with the same instrumental setup as \elf,
and a very high resolution spectrum of NGC\,246 (Jenkins priv. comm.; Rauch
\etal in prep.). We also compared the \elf\ spectra with FUSE and HST spectra of various
hydrogen-rich central stars, which we are currently analysing (Traulsen \etal 2005,
Hoffmann \etal 2005).

The unidentified lines likely stem from metals other than CNO, whose line
lists we consider complete. They are probably not unknown iron
lines, because we do not detect the strongest known lines of this species
(Sect.\,\ref{NFSiPSFe-Analysis}). We suggest that the lines stem from highly
ionized light metals like neon or magnesium for which line lists are inaccurate
and incomplete. Some of the unidentified lines are quite strong and the most
prominent one is located at 1270.25~\AA.

\begin{table}
\caption{Summary of the model atoms used in the model atmosphere and line formation calculations.
The numbers in brackets give the individual line numbers summed into superlines
for the heavy metal ions.
\label{levels_tab}}
\small
\begin{center}
\begin{tabular}{l l r r r}
      \hline
      \hline
      \noalign{\smallskip}
element & ion & NLTE levels & lines&\\
  \noalign{\smallskip}
 \hline
      \noalign{\smallskip}
      H  & \scriptsize I          & 5  & 10  \\
         & \mbox{\scriptsize II}  & 1 & -- \\
      \noalign{\smallskip}
      He & \scriptsize I          & 5  & 3  \\
         & \mbox{\scriptsize II}  & 14 & 91 \\
         & \mbox{\scriptsize III} & 1  & --  \\
      \noalign{\smallskip}
      C  & \mbox{\scriptsize III} & 3  & 1  \\
         & \mbox{\scriptsize IV}  & 54 & 295 \\
         & \mbox{\scriptsize V}   & 1  & 0  \\
      \noalign{\smallskip}
      N  & \mbox{\scriptsize IV}  & 3  & 1  \\
         & \mbox{\scriptsize V}   & 54 & 297 \\
         & \mbox{\scriptsize VI}  & 1  & 0  \\
      \noalign{\smallskip}
      O  & \mbox{\scriptsize V}   &  6 &  4 \\
         & \mbox{\scriptsize VI}  & 54 & 291\\
         & \mbox{\scriptsize VII} & 1  & 0  \\
      \noalign{\smallskip}
      F  & \mbox{\scriptsize IV}  &  2 &  0 \\
         & \mbox{\scriptsize V}   & 8  & 9  \\
         & \mbox{\scriptsize VI}  & 6  & 4  \\
         & \mbox{\scriptsize VII} & 2  & 1  \\
         & \mbox{\scriptsize VIII}& 1  & 0  \\
      \noalign{\smallskip}
      Ne & \mbox{\scriptsize V}   & 9  & 4  \\
         & \mbox{\scriptsize VI}  & 8  & 9  \\
         & \mbox{\scriptsize VII} & 10 & 12 \\
         & \mbox{\scriptsize VIII}& 1  & 0  \\
     \noalign{\smallskip}
      P  & \mbox{\scriptsize III}   & 3  & 1  \\
         & \mbox{\scriptsize IV}    & 15 & 9  \\
         & \mbox{\scriptsize V}     & 18 & 12 \\
         & \mbox{\scriptsize VI}    & 1  & 0  \\
     \noalign{\smallskip}
      Si & \mbox{\scriptsize III}   & 6  & 4  \\
         & \mbox{\scriptsize IV}    & 16 & 44  \\
         & \mbox{\scriptsize V}     & 15 & 20   \\
         & \mbox{\scriptsize VI}    & 20 & 36  \\
         & \mbox{\scriptsize VII}   & 1  & 0  \\
     \noalign{\smallskip}
      S  & \mbox{\scriptsize IV}    & 6  & 4  \\
         & \mbox{\scriptsize V}     & 14 & 16 \\
         & \mbox{\scriptsize VI}    & 18 & 44  \\
         & \mbox{\scriptsize VII}   & 1  & 0  \\
      \noalign{\smallskip}
      Fe & \mbox{\scriptsize VI}  & 7 & 25 & (340\,132)\\
         & \mbox{\scriptsize VII} & 7 & 24 & (86\,504)\\
         & \mbox{\scriptsize VIII}& 7 & 27 & (8\,724)\\
         & \mbox{\scriptsize IX}  & 7 & 25 & (36\,843)\\
         & \mbox{\scriptsize X}   & 1 &  0 & \\
           \noalign{\smallskip}
\hline
\normalsize
     \end{tabular}
\normalsize
\end{center}
\end{table}

\section{Model atmospheres and synthetic spectra}
\label{model}

We used the T\"ubingen Model Atmosphere Package \tmap\ (Werner \etal 2003, Rauch
\& Deetjen 2003) to compute plane-parallel non-LTE line blanketed model
atmospheres in radiative and hydrostatic equilibrium. The models include the
most abundant elements: He, C, N, O, Ne. At first, we computed a small grid
varying the elemental abundances as well as \Teff\ and \logg. The comparison of
the resulting line profiles to the HST and FUSE spectra confirmed previously
published values (except for the N abundance). In particular, the
\ion{O}{v}~1371~\AA\ line is a very sensitive \Teff\ indicator and we confirm
our result from an earlier fit to this line taken with HST/FOS (Werner \& Heber
1993), but now with a high precision: 140\,000$\pm$5000~K. In a second step we
adopted these parameters for \Teff, \logg, and abundances of He, C, O, Ne and
included several other elements to determine their abundances. For this, we
performed line formation iterations keeping fixed the atmospheric structure. In
this manner we investigated the line profiles of N, Si, P, S, and Fe. The F
abundance was determined recently (Werner \etal 2005) and this value is adopted
for our final spectrum synthesis calculation.

\subsection{Atomic data}
\label{atomic_data}

Table~\ref{levels_tab} summarizes our model atoms. We used our standard model
atoms for He, Ne, F, Fe that were described in previously cited work.  Other model
atoms are extended versions of previous datasets or designed completely a new. We
briefly summarize these advancements. We used the databases of
NIST\footnote{http://physics.nist.gov/ PhysRefData/ASD/index.html} (National
Institute of Standards and Technology) and
CHIANTI\footnote{http://wwwsolar.nrl.navy.mil/chianti\_direct\_data.html} for
the energy levels and the Opacity Project and Iron Project (TIPTOPbase\footnote{
http://vizier.u-strasbg.fr/topbase/}) for the oscillator strengths.

The model ions of \ion{C}{iv}, \ion{N}{v} and \ion{O}{vi} were updated and
completed. The most important improvement was made for \ion{O}{vi}. The FUSE and
HST spectra of \elf\ and of other hot stars reveal that the positions of many
listed \ion{O}{vi} lines are inaccurate. We are therefore able to improve atomic
data by determining the line positions from observations. In
Table~\ref{tab:OVIwavelengths} we list for all these lines the literature values
and our new observationally derived positions. We find deviations up to
0.6~\AA. In our spectrum synthesis we apply the new line positions.

The model atom for phosphorus is entirely new while for sulfur we use a model
designed by Miksa \etal (2002). The silicon model atom was extended to include
lines from ionisation stages higher than \ion{Si}{iv}.

\begin{figure*}
\begin{center}
\includegraphics[width=0.84\textwidth]{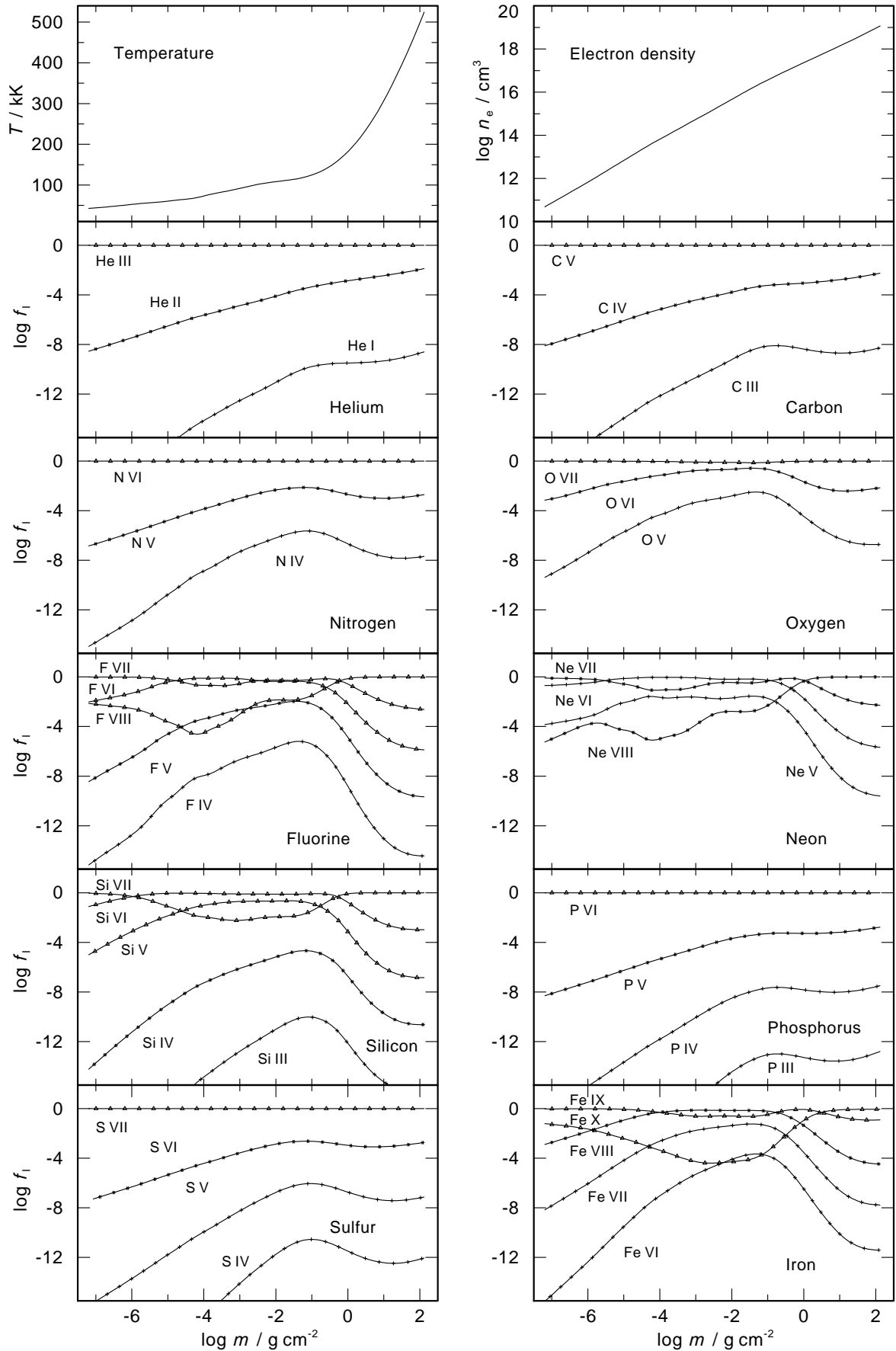}
\caption{Depth dependence of temperature, electron density, and ionisation fractions of all considered
chemical elements in the final atmospheric model (for its parameters see Tab.\,\ref{tab:results}).} 
\label{fig:ionfrac} 
\end{center}
\end{figure*}

Figure~\ref{fig:ionfrac} displays the ionisation stratification of all elements
considered in the final atmospheric model.

\begin{table}
\begin{center}
\caption{Newly determined positions of \ion{O}{vi} lines as used in our improved
  model atom. We compare them to the literature values. The measurements were
  made in the \elf\ spectra described in this paper (J06) as well as in spectra of
  other PG1159 stars (Reiff \etal 2006, R06) and hot central stars (Traulsen
  \etal 2005, T05).}
\label{tab:OVIwavelengths} 
\begin{tabular}{cccc}  \hline \hline 
\noalign{\smallskip}
 Transition &$\lambda_\mathrm{old}$ / \AA     &$\lambda_\mathrm{new}$ / \AA     &  Remark    \\ \hline
\noalign{\smallskip}
$\mathrm{4p\; ^2 P^o_{1/2}} $ - $\mathrm{5s\; ^2 S_{1/2}}   $ &1171.56   &1171.10  & T05     \\
$\mathrm{4p\; ^2 P^o_{3/2}} $ - $\mathrm{5s\; ^2 S_{1/2}}   $ &1172.44   &1172.00  & T05     \\[1.5ex]
                             
$\mathrm{4p\; ^2 P^o_{1/2}} $ - $\mathrm{5d\; ^2 D_{3/2}}   $ &1080.88   &1080.60  & R06        \\
$\mathrm{4p\; ^2 P^o_{3/2}} $ - $\mathrm{5d\; ^2 D_{3/2}}   $ &1081.62   &1081.34  & R06        \\
$\mathrm{4p\; ^2 P^o_{3/2}} $ - $\mathrm{5d\; ^2 D_{5/2}}   $ &1081.52   &1081.24  & R06        \\[1.5ex] 
                             
$\mathrm{4f\; ^2 F^o_{5/2}} $ - $\mathrm{5d\; ^2 D_{3/2}}   $ &1126.46   &1126.28  & R06        \\
$\mathrm{4f\; ^2 F^o_{5/2}} $ - $\mathrm{5d\; ^2 D_{5/2}}   $ &1126.35   &1126.17  & R06        \\
$\mathrm{4f\; ^2 F^o_{7/2}} $ - $\mathrm{5d\; ^2 D_{5/2}}   $ &1126.46   &1126.28  & R06        \\[1.5ex]
                             
$\mathrm{5p\; ^2 P^o_{1/2}} $ - $\mathrm{7d\; ^2 D_{3/2}}   $ &1262.10   &1261.70  &J06  \\
$\mathrm{5p\; ^2 P^o_{3/2}} $ - $\mathrm{7d\; ^2 D_{3/2}}   $ &1262.20   &1261.80  &J06  \\
$\mathrm{5p\; ^2 P^o_{3/2}} $ - $\mathrm{7d\; ^2 D_{5/2}}   $ &1261.82   &1261.42  &J06  \\[1.5ex] 
                                                                                 
$\mathrm{5p\; ^2 P^o_{1/2}} $ - $\mathrm{8d\; ^2 D_{3/2}}   $ &1018.21   &1018.81  &J06  \\
$\mathrm{5p\; ^2 P^o_{3/2}} $ - $\mathrm{8d\; ^2 D_{3/2}}   $ &1018.28   &1018.88  &J06  \\
$\mathrm{5p\; ^2 P^o_{3/2}} $ - $\mathrm{8d\; ^2 D_{5/2}}   $ &1018.15   &1018.75  &J06  \\[1.5ex] 
                             
$\mathrm{5d\; ^2 D_{3/2}}   $ - $\mathrm{7f\; ^2 F^o_{5/2}} $ &1289.82   &1290.06  & T05     \\
$\mathrm{5d\; ^2 D_{5/2}}   $ - $\mathrm{7f\; ^2 F^o_{5/2}} $ &1289.97   &1290.21  & T05     \\
$\mathrm{5d\; ^2 D_{5/2}}   $ - $\mathrm{7f\; ^2 F^o_{7/2}} $ &1289.94   &1290.18  & T05     \\[1.5ex] 
                             
$\mathrm{5f\; ^2 F^o_{5/2}} $ - $\mathrm{7d\; ^2 D_{3/2}}   $ &1293.14   &1293.04  & T05     \\
$\mathrm{5f\; ^2 F^o_{5/2}} $ - $\mathrm{7d\; ^2 D_{5/2}}   $ &1292.74   &1292.64  & T05     \\
$\mathrm{5f\; ^2 F^o_{7/2}} $ - $\mathrm{7d\; ^2 D_{5/2}}   $ &1292.77   &1292.67  & T05     \\[1.5ex] 
                             
$\mathrm{5f\; ^2 F^o_{5/2}} $ - $\mathrm{7g\; ^2 G_{7/2}}   $ &1291.91   &1291.81  & T05     \\
$\mathrm{5f\; ^2 F^o_{7/2}} $ - $\mathrm{7g\; ^2 G_{7/2}}   $ &1291.94   &1291.84  & T05     \\
$\mathrm{5f\; ^2 F^o_{7/2}} $ - $\mathrm{7g\; ^2 G_{9/2}}   $ &1291.91   &1291.81  & T05     \\[1.5ex]
                             
$\mathrm{5g\; ^2 G_{7/2}}   $ - $\mathrm{7f\; ^2 F^o_{5/2}} $ &1292.12   &1292.02  & T05     \\
$\mathrm{5g\; ^2 G_{7/2}}   $ - $\mathrm{7f\; ^2 F^o_{7/2}} $ &1292.09   &1291.99  & T05     \\
$\mathrm{5g\; ^2 G_{9/2}}   $ - $\mathrm{7f\; ^2 F^o_{7/2}} $ &1292.12   &1292.02  & T05     \\[1.5ex]
                             
$\mathrm{5g\; ^2 G_{7/2}}   $ - $\mathrm{7h\; ^2 H^o_{9/2}} $ &1291.99   &1291.89  & T05     \\
$\mathrm{5g\; ^2 G_{9/2}}   $ - $\mathrm{7h\; ^2 H^o_{9/2}} $ &1292.02   &1291.92  & T05     \\
\vspace*{2pt}                
$\mathrm{5g\; ^2 G_{9/2}}   $ - $\mathrm{7h\; ^2 H^o_{11/2}}$ &1292.02   &1291.92  & T05     \\ \hline 
\end{tabular}  
\end{center}
\end{table}

\section{Spectral analysis}
\label{analysis}

\subsection{Abundance determination of N, Si, P, S, and Fe}
\label{NFSiPSFe-Analysis}

\begin{figure}
\begin{center}
\includegraphics[width=0.48\textwidth]{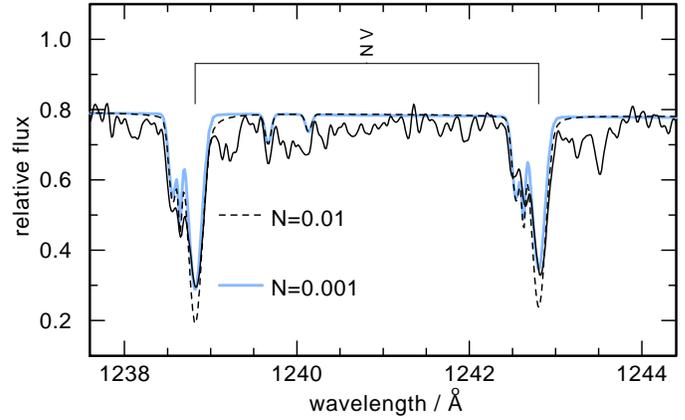}
\caption{The \ion{N}{v} resonance doublet compared with two models with
  different nitrogen abundance. The model with N=0.001 (mass fraction) fits the observed
  profiles very well. Note the weak blueshifted ISM components which are
  included in the model.} 
\label{fig:N-Vgl} 
\end{center}
\end{figure}

\emph{Nitrogen.---}\,Our HST spectrum allows us to distinguish the interstellar and
photospheric components of the \ion{N}{v} resonance doublet
(Fig.\,\ref{fig:N-Vgl}). We find a good fit with N=0.001 (mass fraction). This
is one dex lower than the value derived by Dreizler \& Heber (1998) from the
HST/FOS spectrum that could not separate the interstellar and photospheric
components.

\begin{figure}
\begin{center}
\includegraphics[width=0.48\textwidth]{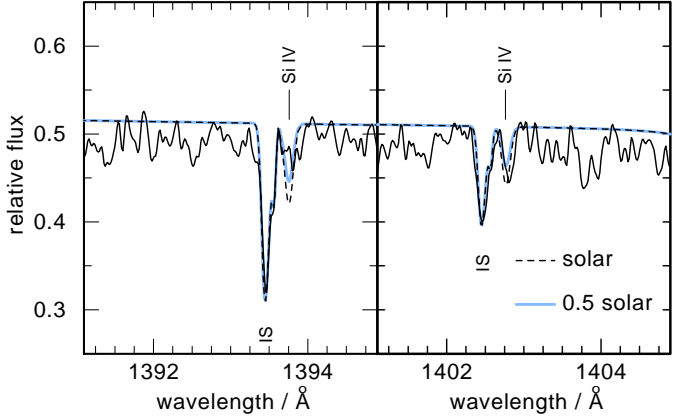}
\caption{The photospheric \ion{Si}{iv} resonance doublet is barely detectable
  while the ISM component (included in the model) is relatively strong. The two model
  profiles suggest that the silicon abundance is about 0.5 solar.} 
\label{fig:Si-Vgl} 
\end{center}
\end{figure}

\begin{figure}
\begin{center}
\includegraphics[width=0.40\textwidth]{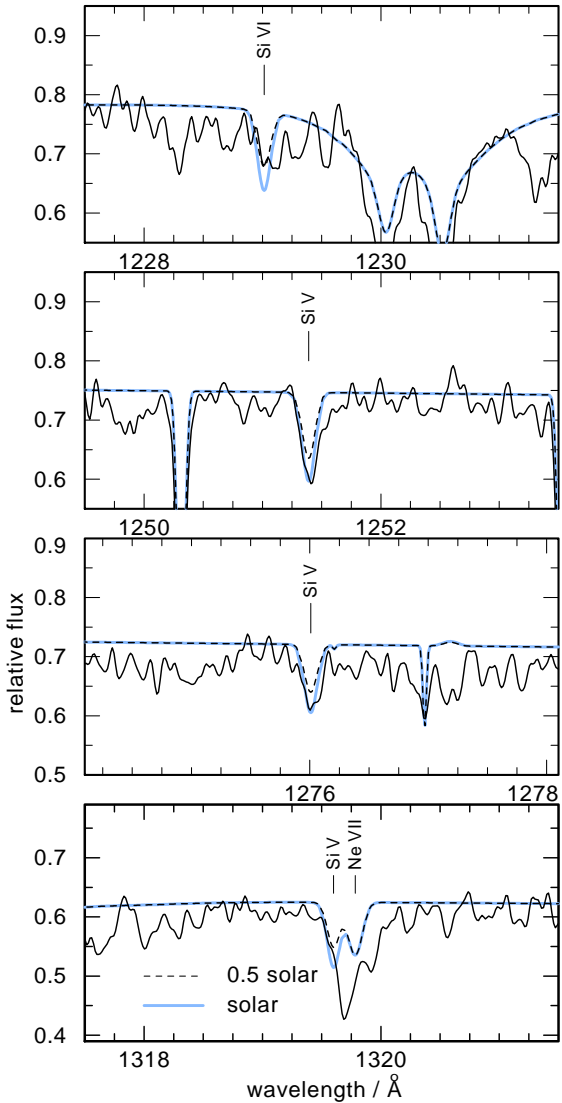}
\caption{Examples for newly discovered lines from highly ionised silicon
  (\ion{Si}{v-vi}) compared with models with solar and 0.5 solar Si abundance.} 
\label{fig:Si-Vgl1} 
\end{center}
\end{figure}

\emph{Silicon.---}\,Figures~\ref{fig:Si-Vgl} and \ref{fig:Si-Vgl1} show the
resonance doublet of \ion{Si}{iv} as well as lines from \ion{Si}{v-vi} together
with two synthetic spectra with solar and half-solar Si abundance. We conclude
that the Si abundance is about 0.5 solar.

\begin{figure}
\begin{center}
\includegraphics[width=0.48\textwidth]{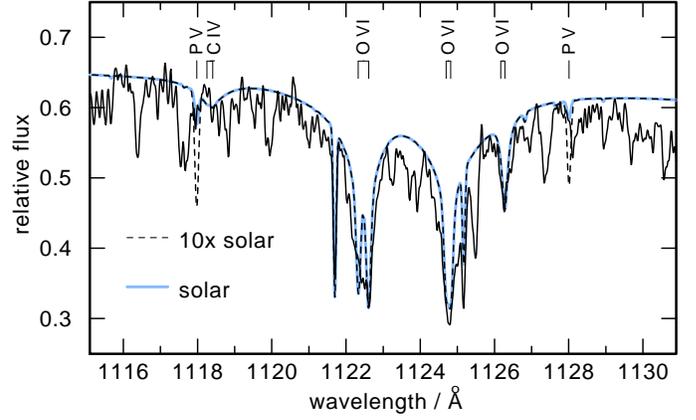}
\caption{The non-detection of the \ion{P}{v} resonance doublet and the two model
  spectra suggest that the P abundance is at most solar.} 
\label{fig:P-Vgl} 
\end{center}
\end{figure}

\emph{Phosphorus.---}\,Figure~\ref{fig:P-Vgl} shows the FUSE spectral region
where the resonance doublet of \ion{P}{v} is located. The doublet is not
detected. By comparison with models having solar and ten times solar P
abundance we conclude that the P abundance is at most solar.

\begin{figure}
\begin{center}
\includegraphics[width=0.48\textwidth]{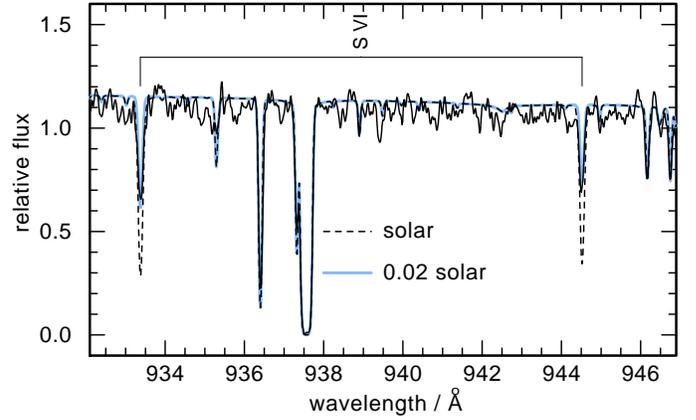}
\caption{The relative weakness of the \ion{S}{vi} resonance doublet suggests a
  strongly subsolar sulfur abundance.
} 
\label{fig:S-Vgl} 
\end{center}
\end{figure}

\emph{Sulfur.---}\,Figure~\ref{fig:S-Vgl} shows the resonance doublet of
\ion{S}{vi} together with two model calculations with solar and 2\% solar
abundance. It is obvious that the S abundance is strongly depleted. We find a
best fit at 2\% solar.

\begin{figure}
\begin{center}
\includegraphics[width=0.40\textwidth]{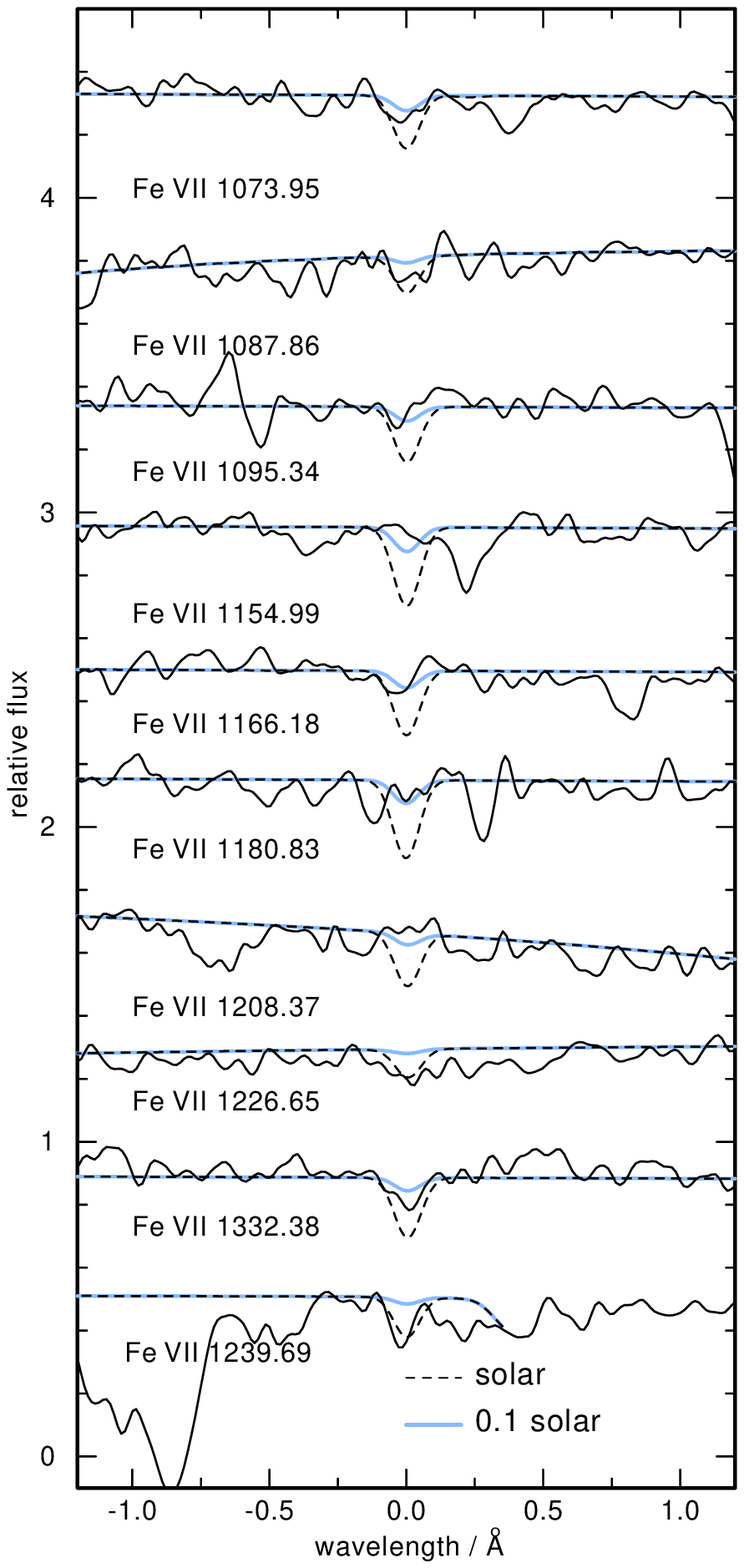}
\caption{Computed \ion{Fe}{vii} lines (with solar and 0.1 solar Fe abundance)
  compared with observations. There is no clear evidence for the presence of iron
  lines in \elf. This suggests that iron is depleted and we estimate an upper abundance limit of
  0.3 solar.} 
\label{fig:Fe-Vgl} 
\end{center}
\end{figure}

\emph{Iron.---}\,We have closely inspected the HST and FUSE spectra looking for
iron lines. The only Fe lines predicted from our models stem from
\ion{Fe}{vii}. The dominant ionisation stages are
\ion{Fe}{viii-ix} (Fig.\,\ref{fig:ionfrac}) but no detectable lines with
accurately known wavelength position are known from this ion (i.e.\ no such
lines are listed in the respective Kurucz (1991) POS files in the relevant
wavelength region). The \ion{Fe}{vii} lines are too weak and we are not able to
make a clear identification. From a comparison of the \elf\ spectrum with those
of (cooler) H-rich central stars and with our model we found that the
\ion{Fe}{vii} line at 1332.38~\AA\ should be among the strongest. There
is an absorption feature in the HST spectrum at this location
(Fig.\,\ref{fig:Fe-Vgl}), but the case is less convincing with other lines. We
cannot detect beyond doubt any iron line. The computed line
profiles from models with solar and 0.1 solar Fe abundance suggest an upper
limit of Fe$\approx$0.3 solar.

We have also tried to look for lines from nickel, which is an important element
in the context of the Fe depletion discovered in \elf\ as well as in other PG1159
stars. The situation for Ni is even less certain than for Fe. The highest
ionisation stage having precisely known line positions in the HST and FUSE
ranges is \ion{Ni}{vi}.  The ionisation structure of Ni in the atmosphere is
comparable to Fe and our models predict no detectable \ion{Ni}{vi} lines at all
because of the very high effective temperature.

\subsection{Absorption by interstellar material}

Absorption by interstellar gas was modelled to facilitate analysis of the
photospheric spectrum.   Species detected in the data and included in the model
are: \ion{H}{i}, \ion{D}{i}, \ion{C}{i}, \ion{C}{ii}, \ion{C}{ii}*,
\ion{C}{iii},  \ion{C}{iv}, \ion{N}{i}, \ion{N}{ii}, \ion{N}{iii}, \ion{N}{v},
\ion{O}{i},  \ion{Mg}{ii}, \ion{Al}{ii}, \ion{Si}{ii}, \ion{Si}{iii},
\ion{Si}{iv},  \ion{S}{ii}, \ion{S}{iii}, \ion{S}{iv}, \ion{Ar}{i},
\ion{Mn}{ii},  \ion{Fe}{ii}, and H$_2$ (J=0,1,2,3).   All interstellar features
in the FUSE and STIS spectra were fitted simultaneously.  A satisfactory fit to
all the neutral and singly-ionized species typical of the warm diffuse ISM was
obtained with a single component at a heliocentric velocity of -9.7\,km/s.  The
\ion{C}{i} and H$_2$ features were fit with a separate component; the Doppler
width was much smaller than that of the warm gas, as expected, but the radial
velocity was the same.  The \ion{C}{iii}, \ion{Si}{iii}, \ion{S}{iii} and
\ion{S}{iv} absorption features were also at about the  same velocity as the
low-ionization species, with some scatter of a few km/s. The highly-ionized
species of \ion{Si}{iv}, \ion{C}{iv}, \ion{N}{v}, and \ion{O}{vi} exhibit a more extended
distribution in velocity.  The \ion{N}{v} profiles show two distinct components,
one at $-12$~km/s and one at $+14$~km/s.  The \ion{Si}{iv}, \ion{C}{iv}, and \ion{O}{vi} 
profiles span the same range of velocity, but the substructure is not as clearly resolved.
These velocities range from a value similar to that of the ISM absorbers to one that
is intermediate between the ISM material and the photosphere.  This suggests that
the highly-ionized material is circumstellar; a detailed analysis of this material
is outside the scope of this paper.
The final ISM model was evaluated for each pixel in the observed
spectra, and multiplied by the photospheric model to produce the plots shown in
Figures\,\ref{fig:FUSE} and \ref{fig:STIS}.

We derived the neutral H column density towards \elf\ by a fit to the
interstellar Ly$_{\alpha}$ line (Fig.\,\ref{fig:STIS}).  Prior to fitting, the
spectrum was normalized by the photospheric model  to account for the broad
\ion{He}{ii} and \ion{C}{iv} absorption features; consideration of the
photospheric \ion{C}{iv} lines at 1210.61~\AA\ (4s--7p) and 1213.56~\AA\
(4d--8f) is particularly important when attempting to derive $n_{\rm H}$ from
Ly$_{\alpha}$ for stars in this spectral class.  Uncertainty in the proper
background level was assessed by performing a series of fits to the
Ly$_{\alpha}$ profile in which the residual background level was varied from
-0.06 to +0.04, in steps of 0.02.  The best fit was obtained with a background
level of 0.0 and  $n_{\rm H} = 1.50 \cdot 10^{20}$~cm$^{-2}$.  The mean
background level in the central portion of the Ly$_{\alpha}$ trough, apart from
the narrow geocoronal emission feature, is $\approx$-0.02.  For this value of
the background, the best-fit $n_{\rm H}$ is  1.45$\cdot 10^{20}$~cm$^{-2}$; we
therefore take 0.05$\cdot 10^{20}$ as representative of the uncertainty in the
determination of $n_{\rm H}$ arising from the uncertainty in the background
determination.

The effect of the uncertainty in \Teff\ was assessed by repeating the series of
fits on spectra that had been normalized by models with  \Teff=136\,000\,K and
146\,000\,K.  The resulting chi-squared contours had the same minimum position
and same gradients away from the minimum, so uncertainty in \Teff\ does not
appear to contribute significantly to the overall uncertainty in $n_{\rm H}$.

Our adopted value of $n_{\rm H}=1.50 \pm 0.05 \cdot 10^{20}$~cm$^{-2}$ is fairly
consistent with earlier measurements ($n_{\rm H}=1.0 \cdot 10^{20}$~cm$^{-2}$,
Kruk \& Werner 1998).  We found that it is not necessary to apply interstellar
reddening to our model in order to reproduce the continuum shape of the observed
spectrum. This is also in accordance with what we found from the HUT data (Kruk
\& Werner 1998).

\subsection{Summary of spectral analysis}

Table~\ref{tab:results} summarizes the results of our spectral analysis
including results from earlier works. We also list mass, luminosity, and
distance estimates taken from Werner \etal (1991).

The overall fit of our final model to the FUSE and HST spectra is good
(Figs.\,\ref{fig:FUSE} and \ref{fig:STIS}). Let us comment on those few
photospheric features where deviations between model and observations are
obvious. The \ion{C}{iv} 4p--10d line at 1060~\AA\ is not seen in the observed
spectrum and is too strong in the model. The explanation could be that the upper
level population is not computed realistically enough because the $n=10$ levels
are the highest ones treated in NLTE. The $n>10$ levels are treated in LTE which
might be insufficient. The computed profiles of two \ion{O}{vi} lines have no
detectable counterpart in the observed spectrum, namely the 5p--7s and 5d--7p
transitions at 1303.8 and 1302.5~\AA. As we have shown above, the
exact wavelength position of many \ion{O}{vi} transitions, all of them involving
the $n=5$ levels (see Table~\ref{tab:OVIwavelengths}), are uncertain within a
few tenths of an \AA. For the two lines in question we could not find plausibly
shifted counterparts in the observed spectrum. It could be that they are blended
with the close-by strong interstellar features.

Concerning the continuum shape, our model overestimates the flux in the
long-wavelength region of the HST spectrum by a few percent when normalized to
the short-wavelength region. This means that the HST spectrum is harder than our
model, which is unphysical. Because of the high \Teff\ our model has a
Rayleigh-Jeans shape and the stellar spectrum cannot be steeper. Hence, there
are problems with the flux calibration of the HST spectrum.  On the compressed
scale of Fig.~\ref{fig:STIS}, the discrepancy is only apparent at the longest
wavelengths where it has grown to nearly 10 percent, but on an expanded scale
the effect appears to begin longward of about 1420~\AA.

\begin{table}
\begin{center}
\caption{Photospheric and stellar parameters of \elf. We list the direct results from
  spectroscopic analyses and derived quantities (mass, luminosity, distance) from comparison with evolutionary
  tracks. Elemental abundances
  are given in mass fractions (2nd column) and relative to solar abundances
  (Grevesse \& Sauval 2001 values; 3rd column).}
\label{tab:results} 
\begin{tabular}{cccc} \hline \hline 
\noalign{\smallskip}
Parameter   & Result                 & Abundances    & Ref.\\
            &                        & (solar units) &     \\
\noalign{\smallskip}
\hline
\noalign{\smallskip}
\Teff / K   & 140\,000 $\pm$5000     &                       & (1), this work \\
\logg (cgs) & 7.0  $\pm$0.5          &                       & (1)           \\
\noalign{\smallskip}
H           & $\le 0.02$             & $\le 0.027$           & (2) \\
He          & 0.33                   & 1.3                   & (3) \\
C           & 0.48                   & 170                   & (3) \\
N           & 0.001                  & 1.2                   & this work \\
O           & 0.17                   & 22                    & (3) \\
F           & $3.2  \cdot 10^{-6}$   & 6.3                   & (3) \\
Ne          & 0.02                   & 11                    & (3) \\
Si          & $3.6 \cdot  10^{-4}$   & 0.5                   & this work \\
P           & $\le 6.4 \cdot 10^{-6}$& $\le 1$               & this work \\
S           & $5.0 \cdot  10^{-6}$   & 0.02                  & this work \\
Fe          & $\le 3.9 \cdot 10^{-4}$& $\le 0.3$             & this work \\
\noalign{\smallskip}
$n_{\rm H}$ / cm$^{-2}$&$1.5 \cdot 10^{20}$&                 & this work \\
E(B-V)      & 0.0                    &                       & this work \\
\noalign{\smallskip}
$M$/M$_\odot$& $0.60^{+.13}_{-.04}$&                          & (1) \\
$L$/L$_\odot$& $2.73^{+.41}_{-.55}$&                          & (1) \\
$d$/kpc    & $0.80^{+.22}_{-.12}$  &                           & (1) \\
\noalign{\smallskip}
\hline
\end{tabular} 
\\References: \\
(1) Werner \etal 1991,
(2) Werner 1996,
(3) Werner \etal 2005
\end{center}
\end{table}

\section{Summary and discussion}
\label{discussion}

We have analysed high-resolution FUSE and HST spectra of \elf\ covering the
wavelength range 912--1730~\AA. They are dominated by lines from highly ionized
carbon and oxygen plus several light metals (Tables \ref{tab:lines-FUSE} and
\ref{tab:lines-STIS}). More than forty relatively strong absorption lines in
the UV spectra remain unidentified (Table~\ref{tab:unidentified photospheric
lines}). They most probably stem from metals between the CNO elements and the
iron group. We discovered that many spectral lines from the \ion{O}{vi} ion have
inaccurately known wavelength positions (up to 0.6~\AA). Our high resolution
spectra allow for a precise wavelength determination ($< 0.05$~\AA) for about 30
lines (Table~\ref{tab:OVIwavelengths}).

The principal aim of this work was the abundance analysis of trace elements
beyond the main atmospheric constituents He, C, and O
(Table~\ref{tab:results}). Since the PG1159 stars display former intershell
matter on their surface as a consequence of a (late) He-shell flash, the
elemental abundances allow us to place constraints on nucleosynthesis and mixing
processes taking place in the AGB phase. We summarize and discuss our
results in brief. For a more general discussion of abundances in PG1159 stars in
this context see Werner \& Herwig (2006).

\emph{Nitrogen.---}For the first time we succeeded resolving the photospheric
and interstellar components of the \ion{N}{v} resonance line, allowing a
reliable abundance determination. We find N=0.001 (mass fraction), which is an
order of magnitude less than the upper limit previously known. We conclude that
\elf\ suffered a late thermal pulse (LTP), i.e., a late He-shell flash during
the previous H-burning post-AGB phase. The flash did not occur on the white
dwarf cooling track (a very late TP, VLTP), because in this case a larger N
abundance is expected. The N abundance in \elf\ is about one tenth of that found
in other GW~Vir pulsators. Hence, the presence of a high N abundance does not
strictly correlate with the occurrence of pulsations as previously found
(Dreizler \& Heber 1998).

\emph{Fluorine.---}\,The strong F overabundance in \elf\ as well as other PG1159
stars was found and discussed in previous work (Werner \etal 2005). It
emphasizes the important role of AGB stars as effective Galactic fluorine
producers.

\emph{Neon.---}\,The Ne overabundance in \elf\ and other PG1159 stars was also
discussed previously (Werner \etal 2004). It confirms the strong convective
overshoot occurring during He-shell burning in thermally pulsing AGB stars.

\emph{Silicon.---}\,We were able to determine the Si abundance in \elf\ for the
first time. This is based on the \ion{Si}{iv} resonance doublet and on several
lines from \ion{Si}{v-vi}. The latter have never been observed before in any
stellar photospheric spectra. Within the error limit Si is solar. This is in
accordance with expectations from stellar evolution models.

\emph{Phosphorus.---}\,We found a solar abundance as an upper limit. Current
evolutionary models, however, predict overabundances of 4--25 times solar.

\emph{Sulfur.---}\,Based on our first discovery of the \ion{S}{vi} resonance
doublet we find an extreme underabundance of S=0.02 solar. This strongly
contradicts evolutionary models that predict only slight depletions (0.6
solar). One might speculate that the S depletion found here is related to the
sulfur abundance anomaly in Type~II (i.e. C-rich) planetary nebulae. Many of
these exhibit an unexpectedly low S/O ratio in contrast to many Type~I PNe and
H~II regions (Henry \etal 2004, 2006). It was argued, however, that this anomaly
in PNe is not real but due to the use of a model-dependent (and for some unknown
reason incorrect) ionisation correction factor.

\emph{Iron.---}\,We determine an upper limit of Fe=0.3 solar. This is in
agreement with results for three other PG1159 stars where we found upper limits
of 0.01--0.1 solar (Miksa \etal 2002). We have suggested that Fe was transformed
into Ni and heavier elements by s-process neutron captures; however, such strong
Fe depletion in the intershell is not predicted by current AGB star models. On
the other hand, this idea is confirmed by the strongly subsolar Fe/Ni ratio
found in Sakurai's VLTP star (Asplund \etal 1999). In the case of \elf\ we are
not able to determine any constraint for its Ni abundance.

While the abundances of the main atmospheric constituents of PG1159 stars can be
explained quantitatively by evolutionary models for AGB stars that later go
through a (late) thermal pulse, a close look at trace elements abundances
reveals that many details remain unclear. The extreme depletions of sulfur and
iron in \elf\ are most unexpected.

\begin{acknowledgements}
Fits to interstellar absorption features were performed using the 
program Owens.f, developed by Martin Lemoine and the FUSE French Team.
E.R. and T.R. are supported by DFG (grant We 1312/30-1) and DLR (grant 50\,OR\,0201),
respectively. J.W.K. is supported by the FUSE project, funded by NASA contract
NAS5-32985. This work was funded in part under the auspices of the U.S.\
Dept.\ of Energy under the ASC program and the LDRD program (20060357ER) at Los
Alamos National Laboratory (F.H.). 
\end{acknowledgements}


\begin{thebibliography}{}

\bibitem[]{asp99} Asplund, M., Lambert, D.~L., Kipper, T., Pollaco, D., \&
  Shetrone, M.~D. 1999, A\&A, 343, 507

\bibitem[]{dh98} Dreizler, S., \& Heber, U. 1998, A\&A, 334, 618

\bibitem[]{f95} Feibelman, W. A. 1995, PASP, 107, 531

\bibitem[]{f99} Feibelman, W. A. 1999, PASP, 111, 221

\bibitem[]{ga05} Gautschy, A., Althaus, L.~G., \& Saio, H. 2005, \aap, 438, 1013

\bibitem[]{Green1986} Green, R.~F., Schmidt, M., \& Liebert, J.\ 1986, \apjs, 61, 305 

\bibitem[]{Grevesse2001} Grevesse, N., \& Sauval, A.~J.\ 2001, Encyclopedia
of Astronomy and Astrophysics, IOP Publishing Ltd. and Nature Publishing Group, p.\,2453

\bibitem[]{Henry04} Henry, R.~B.~C., Kwitter, K.~B., \& Balick, B. 2004, AJ, 127, 2284

\bibitem[]{Henry06} Henry, R.~B.~C., Skinner, J.~N., Kwitter, K.~B., \& Milingo,
  J.~B. 2006, in Planetary Nebulae in our Galaxy and Beyond, eds.\ M.~J. Barlow,
  R.~H. Mendez, IAU Symp. 234, in press (astro-ph/0605033)

\bibitem[]{ho05} Hoffmann, A. I. D., Traulsen, I., Werner, K., Rauch, T.,
  Dreizler, S., \& Kruk, J.~W. 2005, in 14$^{\rm th}$ European Workshop on White
  Dwarfs, eds.\ D. Koester, S. Moehler, ASP Conference Series, 334, 321

\bibitem[]{hol98} Holberg, J.~B., Barstow, M.~A., \& Sion, E.~M. 1998, ApJS, 119, 207

\bibitem[]{kawaler:94}{Kawaler}, S.~D. \& {Bradley}, P.~A. 1994, ApJ, 427, 415

\bibitem[]{KW98} Kruk, J.~W., \& Werner, K. 1998, ApJ, 502, 858

\bibitem[]{w} Kurucz, R.~L. 1991, in Stellar Atmospheres: Beyond Classical Models,
           eds.\ L.~Crivellari, I.~Hubeny, D.~G.~Hummer, NATO ASI Series C,
           Vol.\ 341, p.\,441

\bibitem[]{liebert:89} Liebert, J., Wesemael, F., Husfeld, D., Wehrse, R., Starrfield, S.~G., \& Sion, E.~M. 1989, AJ, 97, 1440

\bibitem[]{mi02} Miksa, S., Deetjen, J.~L., Dreizler, S., Kruk, J., Rauch, T., \& Werner, K. 2002, A\&A, 389, 953

\bibitem[]{Ra03} Rauch, T., \& Deetjen, J.~L. 2003, in: Stellar Atmosphere Modeling,
        eds. I. Hubeny, D. Mihalas, K. Werner, ASP Conference Series, 288, 103

\item Reiff, E., Jahn, D., Rauch, T., Werner, K., Kruk, J.~W., \& Herwig,
  F. 2006, in Astrophysics of Variable Stars, eds. C. Sterken, C. Aerts, ASP Conference Series, 349, 323

\bibitem[]{sahnow:00} Sahnow, D.~J., Moos, H.~W., Ake, T.~B., \etal 2000, ApJL, 538, 7

\bibitem[]{starrfield:83} Starrfield, S., Cox, A.~N., Hodson, S.~W., \& Pesnell, W.~D. 1983, ApJ Lett., 268, 27

\bibitem[]{tra05} Traulsen, I., Hoffmann, A. I. D., Werner, K., Rauch, T.,
  Dreizler, S., \& Kruk,
  J.~W. 2005, in 14$^{\rm th}$ European Workshop on White Dwarfs, eds.\ D. Koester,
  S.~Moehler, ASP Conference Series, 334, 325

\bibitem[]{w96} Werner, K. 1996, A\&A, 309, 861

\bibitem[]{wh93} Werner, K., \& Heber, U. 1993, in 
	White Dwarfs: Advances in Observation and Theory, NATO ASI Series C,
	Vol.\ 403, ed.\ M.A.\ Barstow, Kluwer, Dordrecht,  p. 303

\bibitem[]{werner:06} Werner, K., \& Herwig, F. 2006, PASP, 118, 183

\bibitem[]{w91} Werner, K., Heber, U., \& Hunger, K. 1991, A\&A, 244, 437

\bibitem[]{wetal96} Werner, K., Dreizler, S., Heber, U., Rauch, T., Fleming, T.~A., Sion,
	E.~M., \& Vauclair, G. 1996, A\&A, 307, 860

\bibitem[]{w03}  Werner, K., Dreizler, S., Deetjen, J.~L., Nagel, T., Rauch, T., \& Schuh, S.~L. 2003,
        in: Stellar Atmosphere Modeling, eds. I. Hubeny, D. Mihalas, K. Werner, ASP Conference Series, 288, 31
 
\bibitem[]{w04} Werner, K., Rauch, T., Reiff, E., Kruk, J.~W., \& Napiwotzki, R.\ 2004, \aap, 427, 685 

\bibitem[]{w05} Werner, K., Rauch, T., \& Kruk, J.~W.\ 2005, \aap, 433, 641 

\bibitem[]{Wesemael1985} Wesemael, F., Green, R.~F., \& Liebert, J.\ 1985, \apjs, 58, 379 
  
\end{thebibliography}
\end{document}